\documentclass[12pt]{article}
\usepackage{graphicx}

\newcommand\pubnumber{SNSN-323-63}
\newcommand\pubdate{\today}

\def\napoli{Department of Physics and Technology\\
University of Bergen, 5007 Bergen, NORWAY}
\def\support{\footnote{Work supported by the Norwegian Research Council.}}

\def\Title#1{\begin{center} {\Large #1 } \end{center}}
\def\Author#1{\begin{center}{ \sc #1} \end{center}}
\def\Address#1{\begin{center}{ \it #1} \end{center}}

\newcommand\pubblock{\rightline{\begin{tabular}{l} \pubnumber\\
         \pubdate  \end{tabular}}}
\newenvironment{Abstract}{\begin{quotation}  }{\end{quotation}}
\newenvironment{Presented}{\begin{quotation} \begin{center} 
             PRESENTED AT\end{center}\bigskip 
      \begin{center}\begin{large}}{\end{large}\end{center} \end{quotation}}
\def\Acknowledgements{\bigskip  \bigskip \begin{center} \begin{large}
             \bf ACKNOWLEDGEMENTS \end{large}\end{center}}

\def\beq{\begin{equation}}
\def\eeq#1{\label{#1}\end{equation}}
\def\eeqn{\end{equation}}

\def\beqa{\begin{eqnarray}}
\def\eeqa#1{\label{#1}\end{eqnarray}}
\def\eeqan{\end{eqnarray}}

\let\bar=\overbar

\def\Dslash{\not{\hbox{\kern-4pt $D$}}}
\def\dslash{\not{\hbox{\kern-2pt $\del$}}}

\def\BR{\mbox{\rm BR}}

\def\msb{{\bar{\ssstyle M \kern -1pt S}}}

\input babarsym.tex
\def\CP{$ C \! P$ }

\def\ra{\rightarrow}

\def\babar{\mbox{\slshape B\kern-0.1em{\smaller A}\kern-0.1em B\kern-0.1em{\smaller A\kern-0.2em R}} } 
\def\babarc{\mbox{\slshape B\kern-0.1em{\smaller A}\kern-0.1em B\kern-0.1em{\smaller A\kern-0.2em R}}, } 

\def\ra{\rightarrow}
\begin{document}
\begin{titlepage}
\pubblock

\vfill
\Title{$B \ra K^{(*)} \ell^+ \ell^- $ from B-factories and Tevatron}
\vfill
\Author{Gerald Eigen \\
representing the \babar collaboration  \support}
\Address{\napoli}
\vfill
\begin{Abstract}

\babar and Belle measurements of branching fractions, rate asymmetries and angular observables in the decay modes $B \ra K^{(*)} \ell^+ \ell^-$ are reviewed and new results from CDF on  $B \ra K^{(*)} \mu^+ \mu^-$ branching fractions and angular observables are discussed. A first search for $B^+ \ra K^+ \tau^+ \tau^-$ is presented.

\end{Abstract}
\vfill
\begin{Presented}
CKM workshop 2010\\
Warwick, UK,  September 06--10, 2010
\end{Presented}
\vfill
\end{titlepage}
\def\thefootnote{\fnsymbol{footnote}}
\setcounter{footnote}{0}

\section{Introduction}
The decays $b \ra s \ell^+ \ell^-$, where $\ell^+ \ell^-$ is an $e^+ e^-, \mu^+ \mu^-$ or $\tau^+ \tau^-$ pair, are flavor-changing neutral current  (FCNC) processes, which are forbidden in the Standard Model (SM) at tree level but are allowed to proceed via electroweak loops and weak box diagrams. An effective Hamiltonian is used to calculate decay amplitudes \cite{SM}, which depend on three effective Wilson coefficients, $C_7^{eff}$, $C_9^{eff}$, and  $C_{10}^{eff}$. The first is extracted from the $B \ra X_s \gamma$ branching fraction, the latter two respectively represent the vector and axial vector part of the weak penguin and box diagrams. New Physics effects involve new loops that interfere with the SM processes modifying the measured values of  $C_7^{eff}$, $C_9^{eff}$, and  $C_{10}^{eff}$ with respect to the SM predictions \cite{NP}. In addition, scalar and pseudoscalar processes may contribute that introduce new Wilson coefficients $C_s$ and $C_p$ that are forbidden in the SM. Thus, it is important to measure many observables in order to overconstrain the complex Wilson coefficients \cite{lee}. These electroweak penguin modes contribute in probing New Physics at a scale of a few TeV \cite{nir}. In this review, we focus on exclusive decays presenting results from \babar , Belle and CDF. The data samples are based on luminosities of $349~ \rm fb^{-1}$, $605~ \rm fb^{-1}$ and $4.4~ \rm fb^{-1}$ corresponding to 384 million $B \bar B$ events, 656 Million $B \bar B$ events and $2 \times 10^{10} ~b \bar b$ events, respectively.

\section{Selection of $B \ra K^{(*)} e^+ e^-$ and $B \ra K^{(*)} \mu^+ \mu^-$ Events}
 \babar and Belle fully reconstruct ten $B \ra K^{(*)} e^+e^-$ and $B \ra K^{(*)} \mu^+\mu^-$  final states, in which a $K^+, K^0_S,
K^+ \pi^-, K^+ \pi^0$ or $K^0_S \pi^+$ recoils against the lepton pair\footnote{Charge conjugation is implied unless otherwise stated.}, while CDF reconstructs  $K^+ \mu^+\mu^-$ and $K^+ \pi^- \mu^+\mu^-$ final states. \babar (Belle) selects lepton candidates with momenta $ p_e >0.3 (0.4)~\rm GeV/c$ and $ p_\mu >0.7 (0.7)~\rm GeV/c$. \babar and Belle require good particle identification (PID) for $e, \mu, K$, and $\pi$, and select $K^0_S$ in the $\pi^+ \pi^-$ channel. CDF requires muons with $p_T (\mu) > 0.4~\rm GeV/c$, kaons and pions with $p_T (K,\pi) > \rm 1~ GeV/c$ and $B$-mesons with $p_T (B) > 6 ~ \rm GeV/c$. Both, muons and hadrons must have good PID and the muon pair must originate from a secondary vertex. All three experiments suppress
combinatorial $B \bar B$ and $q \bar q$ continuum backgrounds ($q=u,d,s,c)$. Here, the leptons dominantly originate from semileptonic $b$ and $c$ decays. \babar trains neural networks (NN) using event shape variables, vertex information, missing energy, and lepton separation near the interaction region (IR) optimized in each mode and each $q^2$ bin\footnote{This is the squared momentum transfer into the dilepton system.}. Belle trains a Fisher discriminant using event shape variables, missing mass, B flavor tagging, and lepton separation in z near the IR. CDF trains NNs using vertex information, the angle between the signed vertex displacement with respect to the B momentum, and the $\mu$ separation. \babar and Belle select signal candidates using the beam-energy substituted mass $m_{ES}=\sqrt{E^{*2}_{beam} - p^{*2}_B}$ and the energy difference $ \Delta E = E^*_B -E^*_{beam}$, where $E^*_{beam}, E^*_B$ and $p^*_B$ are the beam energy, B-meson energy and B-meson momentum in the $\Upsilon(4S)$ center-of-mass frame, respectively. \babar extracts the signal yield from a one-dimensional unbinned extended maximum log-likelihood fit in $m_{ES}$, while  Belle performs a one (two) dimensional unbinned extended maximum log-likelihood fit in $m_{ES}$ (and $m_{K\pi}$) for $K^{(*)} \ell^+ \ell^-$ modes. CDF selects signal candidates from an unbinned maximum log-likelihood fit in the B invariant-mass distribution. All experiments reject events in the $J/\psi$ and $\psi(2S)$ mass regions and require that $K  \mu$ and $K \pi \mu$ masses are not consistent with a $D$ mass to reject background from $B \ra DX$ decays. The rejected charmonium events are used as control samples for various cross checks.

\section{Results for $B \ra K^{(*)} e^+ e^-$ and $B \ra K^{(*)} \mu^+ \mu^-$ Modes}

\begin{figure}[htb]
\centering
\hskip -0.3cm \includegraphics[height=1.6in]{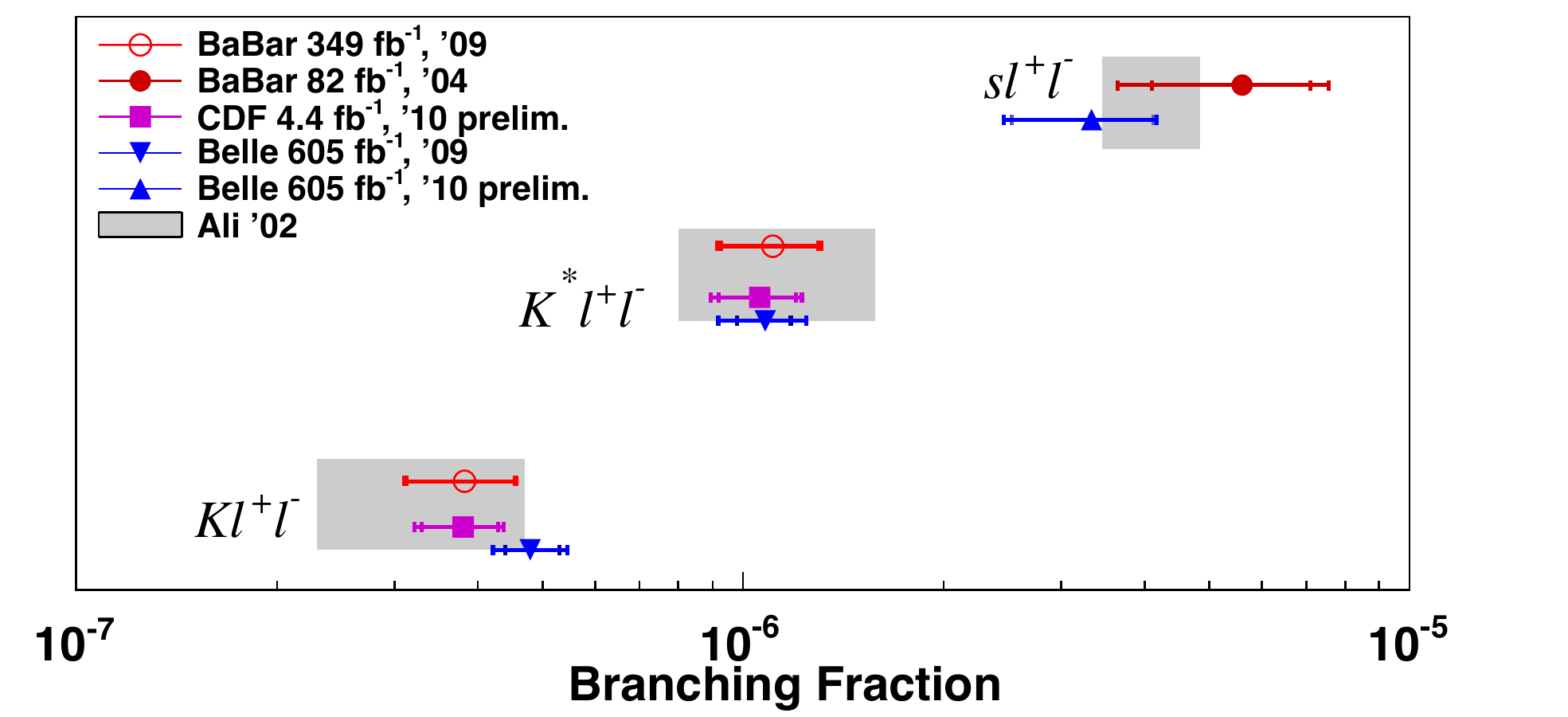} \hskip -0.8cm
\includegraphics[height=1.75in]{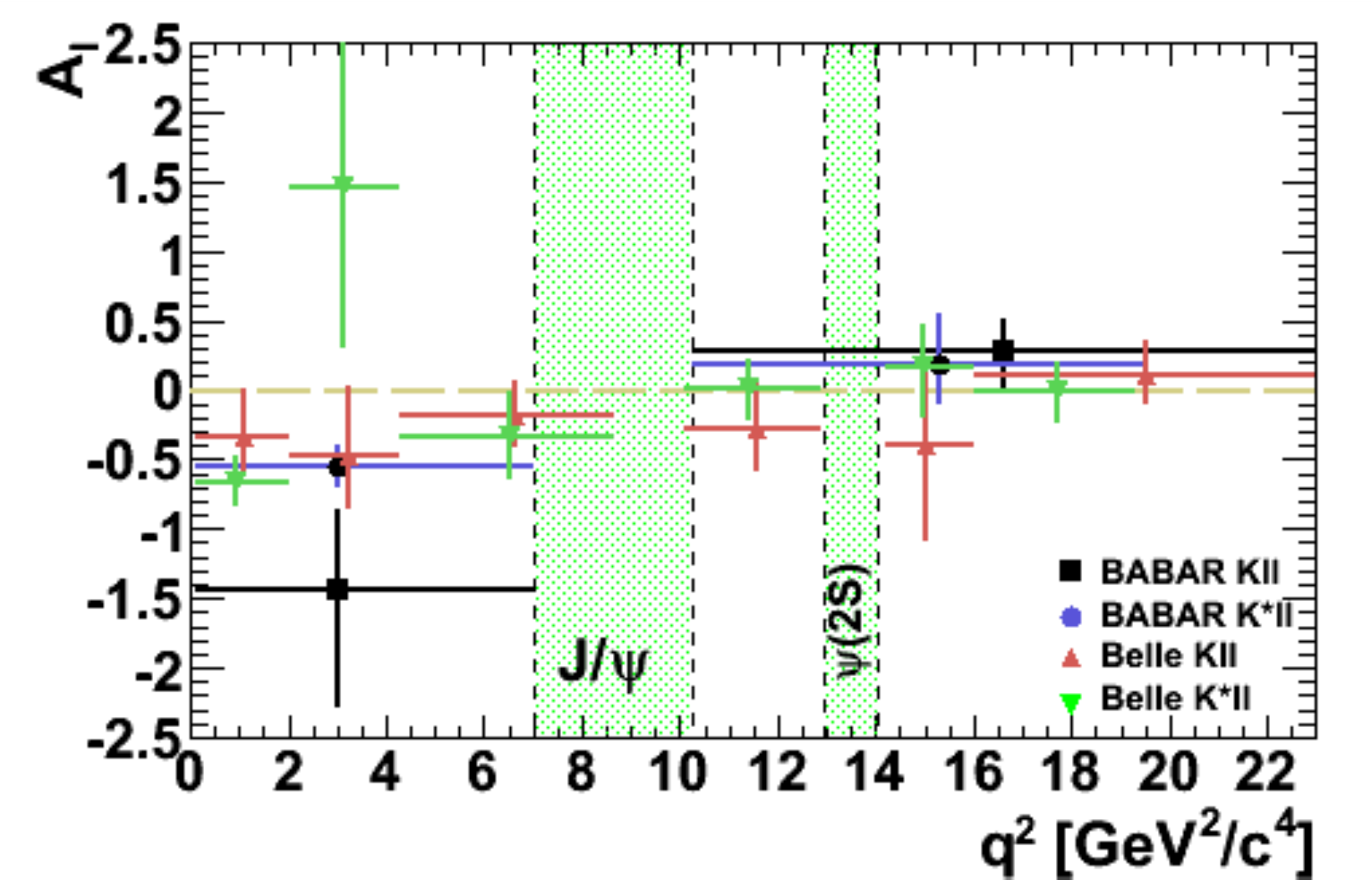}
\caption{(Left) Total branching fractions measurements of $B \ra K^{(*)} \ell^+ \ell^-$ and $B \ra X_s \ell^+ \ell^-$ modes from \babar (red dots), Belle (blue triangles) and CDF (magenta squares) in comparison to the SM prediction (grey-shaded region). For \babar and Belle,  $\ell^+\ell^-$ is a combination of $e^+ e^-$ and $\mu^+ \mu^-$ modes, for CDF it is $\mu^+ \mu^-$. (Right) Isospin asymmetry measurements for $B \ra K^{(*)} \ell^+ \ell^-$  versus $q^2$ from \babar (black squares, blue dots) and Belle (red triangles, green triangles).}
\label{fig:bf}
\end{figure}

Figure~\ref{fig:bf} (left) shows total branching fractions for $B \ra K^{(*)} \ell^+ \ell^-$ ($e^+ e^-$ and $\mu^+ \mu^-$ modes combined) \cite{babar, belle,cdf} and $B \ra X_s \ell^+ \ell^-$\cite{babar2, belle} in comparison to the SM predictions \cite{ali}. The individual exclusive measurements are summarized  in Table~\ref{tab:bf}.  The Belle inclusive measurement is a recent update based on a luminosity of $\rm 605~fb^{-1}$, yielding ${\cal B}(B \ra X_s \ell^+ \ell^-) = 3.33\pm 0.8^{+0.19}_{-0.24}) \times 10^{-6}$ \cite{belle2}. The partial branching fractions measured in the three experiments are also consistent with the SM predictions.
\begin{table}[t]
\begin{center}
\begin{tabular}{|l|c|c|c|c|}  
\hline
Experiment &   Mode &  ${\cal B}~ [10^{-6)}]$ & ${\cal A}_{CP}$ & ${\cal R}_{K^{(*)}}$ \\ \hline
\babar \cite{babar} &   $K \ell^+ \ell^-  $    &  $0.394^{+0.073}_{-0.069}\pm 0.02$  &  $-0.18^{+0.18}_{-0.18}\pm 0.01$ & $0.96^{+0.44}_{-0.34}\pm 0.05$ \\
\babar \cite{babar} &   $K^* \ell^+ \ell^-  $   &   $1.11^{+0.19}_{-0.18}\pm 0.07$  & $-0.01^{+0.16}_{-0.15}\pm 0.01$ &  $ 1.10^{+0.42}_{-0.32}\pm 0.07$ \\ \hline
Belle \cite{belle}  &   $K \ell^+ \ell^-  $     &   $0.48^{+0.05}_{-0.04}\pm 0.03$  & $ 0.04\pm 0.1\pm 0.02$ & $ 1.03\pm 0.19 \pm 0.06$  \\
Belle \cite{belle} &   $K^* \ell^+ \ell^-  $   &    $1.07^{+0.11}_{-0.10}\pm 0.09$ & $-0.10\pm 0.1 \pm 0.01$ &  $0.83 \pm 0.17\pm 0.8$ \\ \hline
CDF \cite{cdf} &   $K \mu^+ \mu^-  $    &    $0.38^{+0.05}_{-0.05}\pm 0.03$  & &  \\ 
CDF  \cite{cdf}&   $K^* \mu^+ \mu^-  $   &   $1.06^{+0.14}_{-0.14}\pm 0.09$ & &     \\ 
\hline
\end{tabular}
\caption{Branching fractions, \CP asymmetries and lepton flavor ratios for $B \ra K^{(*)} \ell^+ \ell^-$ modes in the entire $q^2$ region from \babarc Belle, and CDF. Uncertainties are statistical and systematic, respectively.}
\label{tab:bf}
\end{center}
\end{table}

Rate asymmetries are more precisely measured than branching fractions, since many uncertainties cancel \cite{kruger}. The isospin asymmetry \cite{feldmann02}
\begin{equation}
{\cal A}_I (q^2)= \frac{d{\cal B}(B^0 \ra K^{(*)0} \ell^+ \ell^-)/dq^2 -(\tau_{B^0}/\tau_{B^+}) d {\cal B}(B^+ \ra K^{(*)+} \ell^+ \ell^-)/dq^2}
{d{\cal B}(B^0 \ra K^{(*)0} \ell^+ \ell^-)/dq^2 +(\tau_{B^0}/\tau_{B^+}) d {\cal B}(B^+ \ra K^{(*)+} \ell^+ \ell^-)/dq^2},
\end{equation}
corrected for the different $B^0$ and $B^+$ lifetimes ($\tau_{B^0}/\tau_{B^+}$), 
is expected to be small in the SM ($\d {\cal A}_I (q^2)/dq^2$ is $-0.01$ for $q^2 =2.7-6~\rm GeV^2/c^4$ after dropping from $\simeq 0.075$ at $q^2 =0.1~\rm GeV^2/c^4$ and crossing zero near $q^2=1.7~\rm GeV^2/c^4$) \cite{feldmann02}.  
Figure~\ref{fig:bf} (right) shows the \babar and Belle ${\cal A}_I$ measurements for different $q^2$ regions. The $q^2$ integrated isospin asymmetry and ${\cal A}_I$ for $q^2$ values above the $J/\psi$ are consistent with the SM prediction. Below the $J/\psi$, however, \babar observes a negative ${\cal A}_I$ that deviates significantly from the SM prediction ($ 3.9 \sigma$ from ${\cal A}_I=0$) . For models in which the sign in $C^{eff}_7$  is flipped with respect to the value in the SM, a small negative ${\cal A}_I$ is expected \cite{feldmann02, yan}, but it is too small to explain the \babar measurement. 
For low $q^2$, the Belle results are consistent with both \babar and the SM.

In the SM, the direct \CP asymmetry 
\begin{equation}
{\cal A}_{CP} = \frac{{\cal B}(\bar B\ra K^{(*)} \ell^+ \ell^-) - {\cal B}(B \ra K^{(*)} \ell^+ \ell^-)}
{{\cal B}(\bar B \ra K^{(*)} \ell^+ \ell^-) + {\cal B}(B \ra K^{(*)+} \ell^+ \ell^-)}.
\end{equation}
is expected to be ${\cal O}(10^{-3})$, and new physics at the electroweak scale may provide significant enhancements \cite{bobeth08}.  \babar performs a simultaneous fit to $B^+ \ra K^+ \ell^+ \ell^-$ and $B \ra K^* \ell^+ \ell^-$ modes. The results summarized in Table~\ref{tab:bf} together with Belle's measurements are consistent with the SM expectations.

In the SM, the lepton flavor ratios ${\cal R}_K={\cal B}(B \ra K \mu^+ \mu^-)/ {\cal B}(B \ra K e^+ e^-)$ and  ${\cal R}_{K^*}={\cal B}(B \ra K^* \mu^+ \mu^-)/ {\cal B}(B \ra K^* e^+ e^-)$ integrated over all $q^2$ are predicted to be one and 0.75, respectively. The theoretical uncertainties are just a few percent. For example, in two-Higgs-doublet models the presence of a SUSY Higgs might give $\sim 10\%$ corrections to ${\cal R}_{K^{(*)}}$ for large $\tan \beta$ \cite{yan}.The \babar and Belle measurements summarized in Table~\ref{tab:bf} are consistent with the SM expectations. 

The $B \ra K^* \ell^+ \ell^-$ angular distribution depends on three angles: $\theta_K$, the angle between the K momentum and the B momentum in the $K^*$ rest frame, $\theta_\ell$, the angle between the $\ell^+ (\ell^- )$ momentum and the $B (\bar B) $ momentum in the $\ell^+ \ell^-$ rest frame, and $\phi$, the angle between the two decay planes. The angular distribution involves 12 $q^2$-dependent coefficients $J_i$ \cite{kruger1, kim} that can be extracted from a full angular fit in individual bins of $q^2$. Since large data samples are necessary for this study, \babar, Belle and CDF have analyzed only the one-dimensional angular distributions 
\begin{eqnarray}
W(\cos \theta_K) = \frac{3}{2} {\cal F}_L \cos^2 \theta_K + \frac{3}{4} (1-{\cal F}_L )\sin^2 \theta_K, \ \ \ \ \ \ \ \ \ \ \ \ \ \ \ \ \ \ \ \ \ \\
W(\cos \theta_\ell)= \frac{3}{4} {\cal F}_L \sin^2 \theta_\ell  + \frac{3}{8} (1-{\cal F}_L) (1+\cos^2 \theta_\ell) + {\cal A}_{FB} \cos \theta_\ell,
\end{eqnarray}
\noindent
where ${\cal F}_L $ is the $K^*$ longitudinal polarization and ${\cal A}_{FB}$ is the lepton forward-backward asymmetry. While Belle and CDF measure  ${\cal F}_L $  and ${\cal A}_{FB}$ in six $q^2$ bins, \babar measured ${\cal F}_L $  and ${\cal A}_{FB}$ in two $q^2$ bins due to the limited data sample. An update with the full \babar data set in six $q^2$ bins is in progress. The measured $m_{ES}$ and angular distributions are fitted with signal, combinatorial background and peaking background components. After determining the signal yield from the $m_{ES}$ spectrum, ${\cal F}_L$ is extracted from a fit to the $\cos \theta_K$ distribution for  fixed signal yield. Finally, ${\cal A}_{FB}$ is extracted from the $\cos \theta_\ell$ distribution for fixed signal yield and fixed ${\cal F}_L$. 

\begin{figure}[htb]
\centering
 \hskip -0.3cm \includegraphics[height=1.75in]{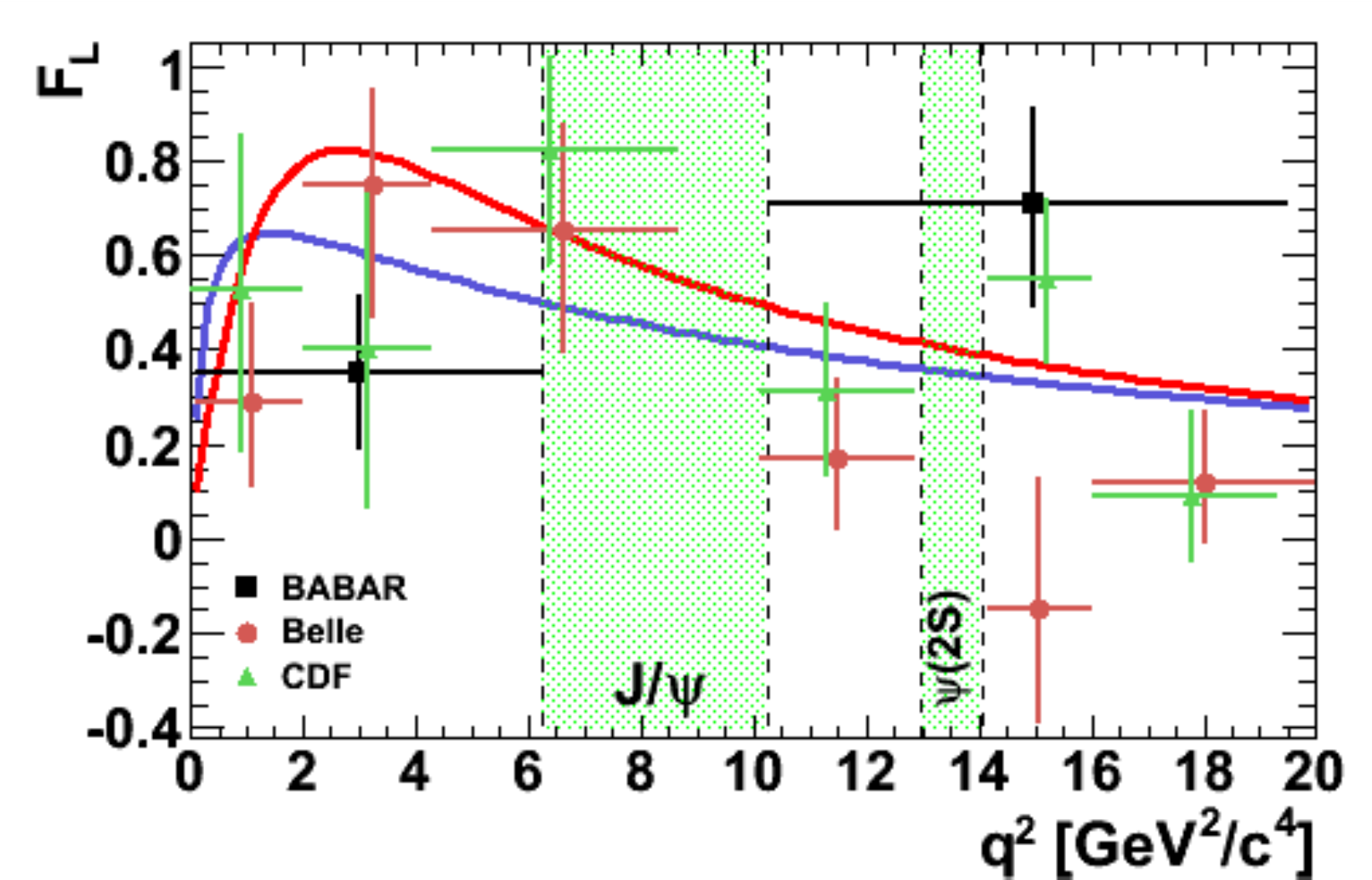} \hskip -0.3cm
\includegraphics[height=1.75in]{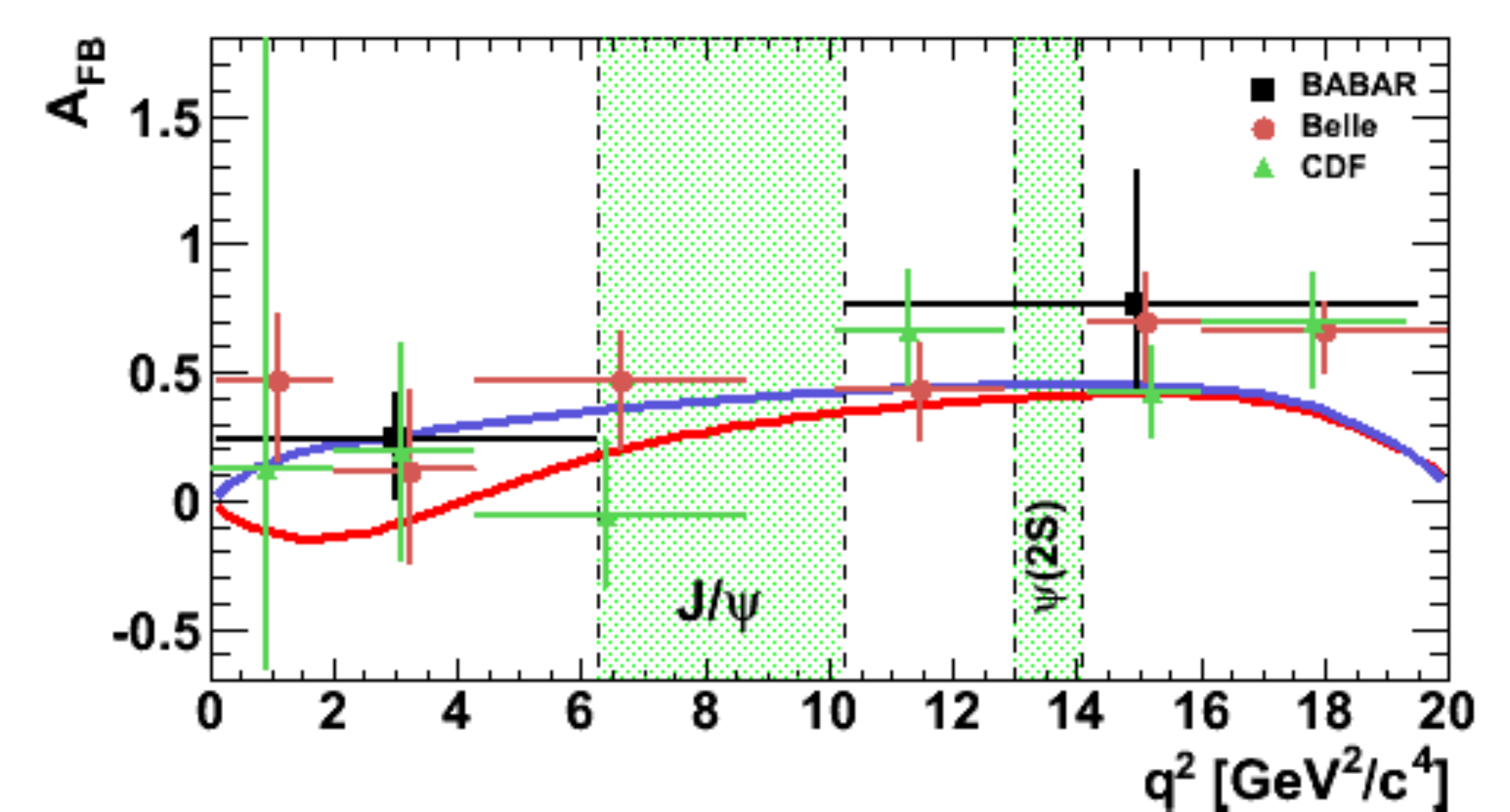}
\caption{(left) Measurements of ${\cal F}_L$ and (right) Measurements of  ${\cal A}_{FB}$ in  $B \ra K^{(*)} \ell^+ \ell^-$ modes by
\babar (black squares), Belle ( brown dots) and CDF (green triangles). The SM prediction (flipped-sign $C^{eff}_7$ model) is shown by the upper red (lower blue) curve for ${\cal F}_L$ and the lower red (upper blue) curve for ${\cal A}_{FB}$.}
\label{fig:afb}
\end{figure}

Figure~\ref{fig:afb} shows the \babarc Belle, and CDF results for ${\cal F}_L$ (left) and ${\cal A}_{FB}$ (right) in comparison to the SM prediction (lower red curve)~\cite{buchalla} and for flipped-sign $C^{eff}_7$ models (upper blue curve) \cite{hou, kruger2}. In the SM, ${\cal A}_{FB}$ is negative for small $q^2$, crosses zero at $q^2_0 =(4.2\pm 0.6)~\rm GeV^2/c^4$ and is positive for large $q^2$, while for flipped-sign $C^{eff}_7$ models ${\cal A}_{FB}$ is positive for all $q^2$. Table~\ref{tab:afb} summarized the ${\cal F}_L$ and ${\cal A}_{FB}$ measurements from $B \ra K^* \ell^+ \ell^-$ in the low $q^2$ region in comparison to the SM prediction. For ${\cal F}_L$, the three measurements are consistent with each other and the SM prediction. For  ${\cal A}_{FB}$, the three measurements are in good agreement. Though they are in better agreement with the flipped-sign $C^{eff}_7$ model, they are consistent with the SM prediction. For $B \ra K \ell^+ \ell^-$,  ${\cal A}_{FB}$ is consistent with zero as expected in the SM. 

\begin{table}[t]
\begin{center}
\begin{tabular}{|l|c|c|c|}
\hline
Experiment & $ q^2$ bin $\rm [GeV^2/c^4]$&  ${\cal F}_L$ & ${\cal A}_{FB}$ \\ \hline  
\babar \cite{babar3} & 0.1-6.25  & $0.35 \pm 0.16 \pm 0.04$  &  $0.24^{+0.18}_{-0.23}\pm 0.05 $ \\
Belle \cite{belle}  &  1-6  & $0.67 \pm 0.23 \pm 0.04$  &  $0.26^{+0.27}_{-0.30}\pm 0.07$\\
CDF  \cite{cdf} &  1-6  &  $0.5^{+0.27}_{-0.30}\pm 0.04$  &   $0.43^{+0.36}_{-0.37}\pm 0.06$ \\  \hline
SM \cite{bobeth} & 1-6 & $0.73^{+0.13}_{-0.23}$ & $-0.05^{+0.03}_{-0.04}$ \\ \hline
\end{tabular}
\caption{\babarc Belle, and CDF measurements of ${\cal F}_L$ and ${\cal A}_{FB}$ from $B \ra K^{*} \ell^+ \ell^-$ modes in the low $q^2$ region.}
\label{tab:afb}
\end{center}
\end{table}

\section{Search for $B^+ \ra K^+ \tau^+ \tau^-$}

In the SM, the $q^2$ dependence of the $B \ra X_s \tau^+ \tau^-$ decay rate has a shape similar to that of $B \ra X_s \mu^+ \mu^-$ in the high $q^2$ region. The  $B^+ \ra K^+ \tau^+ \tau^-$ branching fraction is predicted to be $\sim 2 \times 10^{-7}$ in the SM, which is $50-60\%$ of the total inclusive branching fraction \cite{hewett}. Enhancements are predicted in models beyond the SM. In the next-to-minimal supersymmetric models (NMSSM), for example, the rate may be enhanced by the squared tau-to-muon mass ratio $(m_\tau/m_\mu)^2 \sim 280$. Since signal final states contain 2-4$\nu$, a different analysis strategy is needed here to control backgrounds. 

 \babar  has performed the first  search for $B^+ \ra K^+ \tau^+\tau^-$ using an integrated luminosity of $423~\rm fb^{-1}$ which corresponds to 465 $B \bar B$ events. The recoiling ("tag") $B$ is reconstructed in many hadronic final states, $ B^- \ra D^{(*)0,+}X$, where $X$ represents up to six hadrons ($\pi^\pm, \pi^0, K^\pm, K^0_S$).  Using $m_{ES}$ and $\Delta E$ the tag is selected with an efficiency of $\sim 0.2\%$. The single-prong $\tau$  decays $ \tau \ra e \nu \bar \nu, \tau \ra \mu \nu \bar \nu$ and $\tau \ra \pi \nu$ are selected as signal modes. Thus, signal candidates are required to have only three charged particles of which one is an identified kaon with charge opposite to the tag $B$ and $
 0.44 < p_K < 1.4~\rm GeV/c$ in the center-of-mass frame. The two remaining particles must have opposite charge, be consistent with the signal $\tau$ decays, have $p < 1.59~ \rm GeV/c$ and a mass $ M_{pair} < 2.89~\rm GeV/c^2$. Further requirements are $q^2 = (\vec p_{\Upsilon(4S)} -\vec p_{tag}-\vec p_K )^2 /c^2 > 14.23~ \rm GeV^2/c^4$,  a missing energy ($i.e.$ the energy carried off by neutrinos estimated as the difference between $\Upsilon(4S)$ energy and that of all observed particles) of $1.39 < E_{miss} < 3.38~\rm GeV$, and neutral energy deposited in the electromagnetic calorimeter $E_{extra} < 0.74~ \rm GeV$. Continuum background is suppressed by $| \cos \theta_T | < 0.8$, where $\theta_T$ is the opening angle between the thrust axis of the tag and that of the rest of the event. The largest remaining background originates from $B^+ \ra D^0 X^+$, which is suppressed by combining the signal $K^+$ with the $\tau$ daughter of opposite charge assigned the $\pi$ mass hypothesis and requiring a mass $M_{K\pi} > 1.96~\rm GeV/c^2$.

\babar observes 47 events with an expected background of $64.7\pm7.3$ events.  Including systematic uncertainties a  branching fraction upper limit of ${\cal B}(B \ra K^+ \tau^+ \tau^-) < 3.3 \times 10^{-3}$ is set at $90\%$ confidence level (CL).
 
 
\section{Conclusion}

\babar and Belle have measured branching fractions, rate asymmetries and angular observables in $B \ra K^{(*)} \ell^+ \ell^-$ final states. Recently, CDF contributed new measurements on branching fractions and angular observables in $B \ra K^{(*)} \mu^+ \mu^-$. Except for the isospin asymmetry at low values of $q^2$ all other measurements are consistent with the SM, though ${\cal F}_L$ and ${\cal A}_{FB}$ agree also with the flipped-sign $C^{eff}_7$ model. \babar has performed the first search for $B^+ \ra K^+ \tau^+ \tau^-$ setting a branching fraction upper limit of ${\cal B}(B^+ \ra K^+\tau^+ \tau^-) < 3.3\times 10^{-3}$ at  $90\% ~CL$. Although all experiments are expected to update results with the final data sets, significant improvement in precision will come from LHCb and the Super B-factories. In these new experiments, sufficiently large data samples will be collected to measure the full angular distribution from which the 12 observables $J_i$ \cite{kruger1} can be measured with high precision in different bins of $q^2$. In turn, the Wilson coefficients can be determined with high precision to reveal small discrepancies with respect to the SM predictions \cite{lee, kruger2}.

\Acknowledgements
I would like to thank my \babar colleague K. Flood for useful discussions. This work has been supported by the Norwegian Research Council.


\begin{thebibliography}{99}


\bibitem{SM} G.~Buchalla, A.~J.~Buras and M.~E.~Lautenbacher, Rev.\ Mod.\ Phys.\  {\bf 68}, 1125 (1996);
C. Bobeth, M. Misiak and J. Urban, Nucl. Phys. {\bf B574}, 291 (2000); H.H Asatryan $et~al.$, Phys. Rev. {\bf D65}, 034009 (2002); Phys. Lett. {\bf B507}, 162, (2001); G. Hiller and F.Kr\"uger, Phys.Rev. {\bf D69}, 074020 (2004); M. Beneke, Th. Feldmann, and D. Seidel;  Nucl. Phys.{\bf B612}, 25 (2001);M. Beneke, Th. Feldmann, and D. Seidel;  Eur.Phys.J. {\bf C41}, 173 (2005).
\bibitem{NP} G.~Burdman, Phys.\ Rev.\  {\bf D52}, 6400 (1995);
J.~L.~Hewett and J.~D.~Wells, Phys.\ Rev.\  {\bf D55}, 5549 (1997);
W.~J.~Li, Y.~B.~Dai and C.~S.~Huang, Eur.\ Phys.\ J.\  {\bf C40}, 565 (2005);
Y.~G.~Xu, R.~M.~Wang and Y.~D.~Yang,  Phys.\ Rev.\  {\bf D74}, 114019 (2006);
P.~Colangelo {\it et al.}, Phys.\ Rev.\  {\bf D73}, 115006 (2006);
C.-H.~Chen and C.Q.~Geng, Phys. Rev. D {\bf66} 094018 (2002);C.~Bobeth {\it et al}, Phys. Rev.  {\bf D64} 074014 (2001).
\bibitem{lee} K.S.M. Lee {\it et al.}, Phys. Rev. {\bf D75}, 034016 (2007).
\bibitem{nir} G. Isidori, Y. Nir, G. Prerez, arXiv:1002.0900 (2010).
\bibitem{ali} A.~Ali, E.~Lunghi, C.~Greub and G.~Hiller, Phys.\ Rev.\  {\bf D 66}, 034002 (2002).
\bibitem{babar} B. Aubert {\it et~al.} (\babar collaboration), Phys. Rev. Lett.{\bf  102},  091803 (2009).
\bibitem{belle} J.T. Wei {\it et al.} (Belle collaboration), Phys. Rev. Lett.{\bf 103}, 171801 (2009).
\bibitem{cdf} T. Aaltonen  {\it et al.} (CDF collaboration), CDF note 10047 (2010).
\bibitem{babar2} B. Aubert {\it et~al.} (\babar collaboration), Phys. Rev. Lett.{\bf  93}, 081862 (2004).
\bibitem{belle2} C.C.Chiang (Belle collaboration), talk at ICHEP10 (2010).
\bibitem{kruger} F. Kr\"uger, L. M. Sehgal, N. Sinha and R. Sinha, Phys. Rev. {\bf D61}, 114028 (2000), [Erratum-ibid. {\bf D63}, 019901 (2001)].
\bibitem{feldmann02} T.~Feldmann and J.~Matias, JHEP {\bf 0301}, 074 (2003). 
\bibitem{yan} Q.~S.~Yan, C.~S.~Huang, W.~Liao and S.~H.~Zhu, Phys.\ Rev.\  D {\bf 62}, 094023 (2000).
\bibitem{bobeth08} C. Bobeth, G. Hiller and G. Piranishvili, JHEP {\bf 0807}, 106 (2008).
\bibitem{kruger1} F. Kr\"uger {\it et al.}, Phys. Rev. {\bf D61}, 114028 (2000); Erratum-ibid {\bf D63}, 019901 (2001).
\bibitem{kim} C.S. Kim  {\it et al.}, Phys. Rev. {\bf D 62}, 034013 (2000).
\bibitem{babar3} B. Aubert {\it et~al.} (\babar collaboration), Phys. Rev. {\bf  D79}, 031102 (2009).
\bibitem{buchalla} G. Buchalla $et~al.$, Phys. Rev. {\bf D63}, 014015 (2001). 
\bibitem{buchalla2} G. Buchalla $et~al.$, Phys. Rev. {\bf D63}, 014015 (2001). 
\bibitem{hou} A. Hovhannisyan, W. S. Hou and N. Mahajan, Phys. Rev. {\bf D 77}, 014016 (2008).
\bibitem{hewett} J.L. Hewett, Phys. Rev. {\bf D53}, 4964 (1995).
\bibitem{flood} K. Flood,  talk at the Int. Conf. on HEP, Paris July 22-28 (2010).
\bibitem{kruger2} F. Kr\"uger and J. Matias, Phys. Rev. {\bf D71}, 094009 (2005).
\bibitem{bobeth} C. Bobeth, G. Hiller and D. van Dyk, JHEP {\bf 1007}, 098 (2010). 


\end{thebibliography}
\end{document}